\documentclass[prl,aps,showpacs,twocolumn,superscriptaddress]{revtex4-1}
\usepackage{graphicx}
\usepackage[T1]{fontenc}
\usepackage{amssymb,latexsym,amsthm,amsmath}
\usepackage{listings}
\usepackage{enumerate}
\usepackage{color}

\begin{document}

\title{Inter-layer synchronization in multiplex networks}
\author{R. Sevilla-Escoboza}
\affiliation{Centro Universitario de los Lagos, Universidad de Guadalajara, Jalisco 47460, Mexico}
 \author{I. Sendi\~na-Nadal}
 \affiliation{Complex Systems Group {\& GISC}, Universidad  Rey Juan Carlos, 28933 M\'ostoles, Madrid, Spain}
 \affiliation{Center for Biomedical Technology, Universidad
   Polit\'ecnica de Madrid, 28223 Pozuelo de Alarc\'on, Madrid, Spain}
\author{I. Leyva}
 \affiliation{Complex Systems Group {\& GISC}, Universidad  Rey Juan Carlos, 28933 M\'ostoles, Madrid, Spain}
 \affiliation{Center for Biomedical Technology, Universidad Polit\'ecnica de Madrid, 28223 Pozuelo de Alarc\'on, Madrid, Spain}
\author{R. Guti\'errez}
\affiliation{School of Physics and Astronomy, University of Nottingham, Nottingham, NG7 2RD, UK}
\author{J.M. Buld\'u}
 \affiliation{Complex Systems Group {\& GISC}, Universidad  Rey Juan Carlos, 28933 M\'ostoles, Madrid, Spain}
 \affiliation{Center for Biomedical Technology, Universidad Polit\'ecnica de Madrid, 28223 Pozuelo de Alarc\'on, Madrid, Spain}
 \author{S. Boccaletti}
 \affiliation{CNR- Institute of Complex Systems, Via Madonna del
   Piano, 10, 50019 Sesto Fiorentino, Florence, Italy}
\affiliation{The Italian Embassy in Israel, 25 Hamered st., 68125 Tel Aviv, Israel}

\begin{abstract}
Inter-layer synchronization is a distinctive process of multiplex networks whereby each node in a given layer
undergoes a synchronous evolution with all its replicas in other layers, irrespective of whether or not it is synchronized with the other units of the same layer. We analytically derive the necessary conditions for the existence and stability of inter-layer synchronization, and verify numerically the analytical predictions in several cases where such a state emerges. We inspect the impact of the layer topology on the robustness of such a state against a progressive de-multiplexing of the  network. Finally, we provide experimental evidence by means of multiplexes of nonlinear electronic circuits, showing the stability of the synchronized manifold despite the intrinsic noise and parameter mismatch in the experiment.

\end{abstract}

\maketitle

Synchronization in networked systems is one of the hottest topics of current research in nonlinear science \cite{Boccaletti2006,Arenas2008}.
So far, most of the focus has been concentrated on systems where all the constituents are treated on an equivalent footing, while only in the last
few years the interest has moved towards incorporating the {\it multilayer} character of real world networks, by representing them as graphs formed by diverse layers \cite{kivela2014, Boccaletti2014},
which may either coexist or alternate in time \cite{Boccaletti2014}.
For instance, epidemic processes need a multilayer representation to be properly described \cite{Granell2013}, and also some of the classical examples of pattern formation (like those in Refs.\cite{Winston1991,Marino1999}) find a suitable description within such a formalism  \cite{Asllani2014,Kouvaris2015}.
As far as dynamical processes are concerned, the multilayer formulation allows identifying synchronization regions that arise as a
consequence of the interplay between the layers' topologies \cite{sorrentino2012, *irving2012, Bogojeska2013}, as well as defining new types
of synchronization based on the coordination between layers \cite{Gutierrez2012}.
Several global features have been unveiled: explosive synchronization in multilayer networks \cite{Zhang2015}, synchronization driven by energy transport in interconnected networks \cite{Nicosia2014}, intra-layer
\cite{Gambuzza2014} and cluster \cite{Jalan2014} synchronization in multiplex networks, breathing synchronization in time delayed multiplexes \cite{Louzada2013}, and global synchronization on interconnected layers as in Smart Grids \cite{Bogojeska2013} or in a {\it network of networks} configuration \cite{Aguirre2014}.

In this Letter, we consider multiplex networks, i.e. the case where layers are made of a fixed set of nodes
and connections exist between each node of a layer and all its replicas in the other layers, and show that a novel form of synchronization
emerges, namely {\it inter-layer synchronization}, occurring when each unit in each layer is synchronized with all its replicas, regardless of whether or not it is synchronized with the other members of its layer.
Our results are organized as follows: {\it i)} we analytically derive the conditions for the existence and stability of such a new solution, {\it ii)} we  numerically verify the analytic predictions in several cases where inter-layer synchronization emerges
with or without intra-layer synchronous behaviors, {\it iii)} we  inspect the robustness of the new solution
against a progressive de-multiplexing of the structure, and {\it iv)} we  give experimental evidence of inter-layer synchronization
with nonlinear electronic circuits.

\begin{figure}
  \centering
  \includegraphics[width=\linewidth]{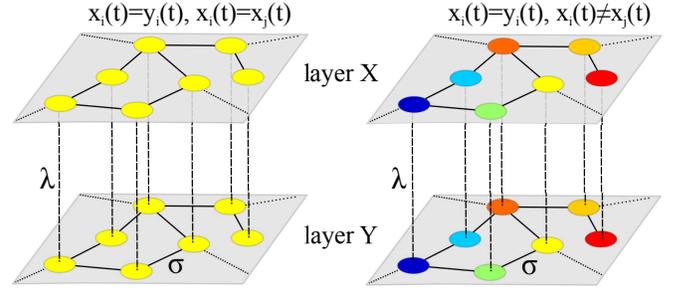}
\caption{(Color online) Schematic representation of a multiplex of two layers of identical oscillators,
and of the two types of inter-layer synchronization: with (left)
  and without (right) intra-layer synchronization. Labels $\sigma$ and $\lambda$ denote the intra- and inter-layer coupling strengths,
respectively. Each node $i$ ($j$) in the top (bottom) layer is an $m$ dimensional dynamical system
 whose state is represented by the vector ${\bf x}_i$ (${\bf y}_j$).}
\label{fig:scheme}
\end{figure}

We start by considering two layers of identical structure,
formed by $N$ identical $m$ dimensional dynamical systems whose states are represented by the vectors ${\bf X}=\{{\bf x}_1,{\bf x}_2,\ldots,{\bf x}_N\}$ (top layer) and ${\bf Y}=\{{\bf y}_1,{\bf y}_2,\ldots,{\bf y}_N\}$ (bottom layer) with ${\bf x}_i, {\bf y}_i, \in \mathbb{R}^m$ for $i = 1,2,\ldots,N$,  as depicted in Fig.~\ref{fig:scheme}.
As already mentioned, the inter-layer synchronous state ${\bf X}\equiv{\bf Y}$
 \footnote{Notice that, when we write here ${\bf X}(t) = {\bf Y}(t)$,
 this has to be understood as a shorthand for the infinitely long time limit $\lim_{t\to\infty} |{\bf X}(t) - {\bf Y}(t)| = 0$.
 Indeed, trajectories starting from different initial conditions cannot become exactly the same for finite times according
 to the theorems on the unicity of the solutions of smooth systems of ordinary differential equations. See, e.g., L. Perko,
 Differential Equations and Dynamical Systems, 3rd edition, Springer-Verlag, New York (2001).}
 can be realized with or without intra-layer synchronization. The former case (Fig.~\ref{fig:scheme} left) corresponds to
 all nodes in both layers following the same trajectory, and it therefore reduces to the classical scenario of a globally synchronous solution whose stability can be accounted for by the Master Stability Function (MSF) \cite{Pecora1998,Boccaletti2006}.
The latter case (Fig.~\ref{fig:scheme} right), instead, is far more general,
as it only requires that every node $i$ in each layer be synchronous to its replica in the other layer [${\bf x}_i(t)={\bf y}_i(t), \forall i$], with unconstrained intra-layer dynamics.


\begin{figure}
  \centering
  \includegraphics[width=\linewidth]{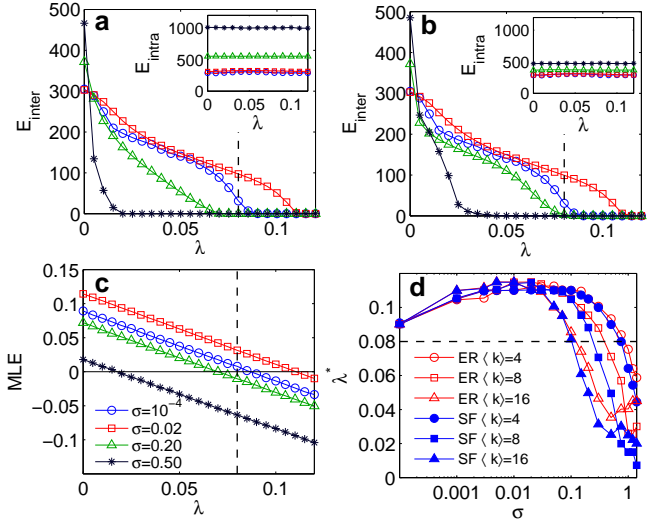}
\caption[]{(Color online).
(a) $E_{inter}$ (see main text for definition) in multiplexes of SF layers of $N=500$ R\"ossler oscillators and (c)
the corresponding MLE as a function of $\lambda$ for several intra-layer couplings $\sigma$
(see legend in panel c). (b) The same as in (a) but for multiplexes of ER layers.
Insets in (a) and (b) show  $E_{intra}$ in the top layer, and the vertical dashed line
is the synchronization coupling threshold for a pair of nodes ($\sigma=0$) coupled through the $y$ variable.
Each point is an average of 10 realizations, with $\langle k\rangle=8$.
(d) Dependence of the inter-layer synchronization onset, $\lambda^*$, on the intra-layer  coupling $\sigma$ for ER (red hollow symbols) and SF (blue solid symbols) layers, and different mean degrees.
The horizontal dashed line is placed at the same value as the vertical line in panels (a)-(c).}
\label{fig2}
\end{figure}

Let the dynamics (in the absence of inter-layer coupling) be $\dot{\bf x}_i = {\bf f}({\bf x}_i) - \sigma \sum_j\mathcal{L}_{ij}\,{\bf h}({\bf x}_j)$ and $\dot{\bf y}_i = {\bf f}({\bf y}_i) - \sigma \sum_j\mathcal{L}_{ij}\,{\bf h}({\bf y}_j)$, where ${\bf f}:  \mathbb{R}^m\rightarrow\mathbb{R}^m$ and ${\bf h}:\mathbb{R}^m\rightarrow\mathbb{R}^m$ are the autonomous evolution and output vectorial functions, $\sigma$ is the intra-layer coupling strength and $\mathcal{L}_{ij}$ are the elements of the Laplacian matrix encoding the intra-layer topology.
In this setting, the layer's dynamical state will be, in general, different at all times, i.e. ${\bf X}(t) \ne {\bf Y}(t)$. Let us now consider the multiplex structure
\begin{eqnarray}
  \label{eq:multiplex}
  \dot {\bf X}= {\bf f}({\bf X}) - \sigma \mathcal{L} \otimes {\bf h}({\bf X}) + \lambda \,[{\bf H}({\bf Y})-{\bf H}({\bf X})],\\
  \dot {\bf Y}= {\bf f}({\bf Y}) - \sigma \mathcal{L} \otimes {\bf h}({\bf Y}) + \lambda \,[{\bf H}({\bf X})-{\bf H}({\bf Y})],
\end{eqnarray}
\noindent
where the inter-layer coupling is realized through the output vectorial function ${\bf H}:\mathbb{R}^m\rightarrow\mathbb{R}^m$ and the inter-layer coupling strength is $\lambda$. Notice that, if the coupling between layers is diffusive, the inter-layer synchronous state {\it always} exists, and the manifold ${\bf X}(t) = {\bf Y}(t)$ is an invariant set whatever value the coupling constants may take.

Let now $\delta {\bf X}(t) = {\bf Y}(t) - {\bf X}(t)$ be the vector describing the difference between the dynamics of the two layers.
Considering a small $\delta {\bf X}$ and expanding around the inter-layer synchronous solution ${\bf Y} = {\bf X} + \delta {\bf X}$ up to first order, one  obtains a set of $N\times m$ linearized equations for the perturbations $\delta {\bf x}_i$:
\begin{equation}\label{eq_variational_id}
\delta \dot{\bf x}_i = \left[ J{\bf f}(\tilde{\bf x}_i) - 2 \lambda\, J{\bf H}(\tilde{\bf x}_i)\right] \delta {\bf x}_i - \sigma \sum_{j} \mathcal{L}_{ij}\, J{\bf h}(\tilde{\bf x}_j)\,\delta {\bf x}_j,
\end{equation}
\noindent where $J$ denotes the Jacobian operator and
${\bf \tilde X}=\left\lbrace {\tilde x_i} \right\rbrace$ is the state of either one isolated layer obeying $\dot{\bf \tilde x}_i = {\bf f}({\bf \tilde x}_i) - \sigma \sum_j\mathcal{L}_{ij}\,{\bf h}({\bf \tilde x}_j)$.
The linear equations (\ref{eq_variational_id}), solved in parallel to  the $N\times m$ nonlinear equations for $\dot{\bf \tilde x}_i$,
allow calculating all Lyapunov exponents transverse to the manifold ${\bf X}= {\bf Y}$. The maximum of those exponents (MLE) as a function of the parameter pair $(\sigma, \lambda)$ actually gives   the necessary conditions for the stability of the inter-layer synchronous solution: whenever MLE$<0$, perturbations transverse to the manifold die out, and the multiplex network is said to be {\it inter-layer synchronizable}.

In the following, the MLE and the inter- and intra-layer synchronization errors ($E_{inter}=\lim_{T\to \infty}\int_0^T\left\lVert \delta {\bf X}(t)\right\rVert dt$ and  $E_{intra}=\lim_{T\to \infty}\int_0^T \sum_{j \neq 1} \left\lVert {\bf x}_j(t)-{\bf x}_1(t)\right\rVert dt$, respectively)
are calculated  by performing numerical simulations of
Eqs.~(\ref{eq_variational_id}). Without lack of generality, we consider two possible kinds of topologies where both layers are either (i) Erd\"os-Renyi \cite{erdos1959} (ER) or (ii) scale-free \cite{Barabasi1999} (SF), in all cases with $N=500$ R\"ossler oscillators \cite{Rossler1976}, whose autonomous evolution is given by ${\bf f}({\bf x})=\left[-y-z,x+0.2 y,0.2+z(x-9.0)\right]$ \footnote{ER and SF networks are generated with the procedures of Refs.~\cite{erdos1959} and \cite{Barabasi1999}, respectively, and therefore the considered SF networks display a degree distribution $p(k) \propto k^{-3}$.}.

We start by setting ${\bf h}=(0,0,z)$ so that the corresponding MSF is in class I \footnote{According to the classification introduced in \cite{Boccaletti2006}, dynamical systems can be classified as: (i) class I, when their MSF is always positive, (ii) classs II, when their MSF is always negative for values higher than a given threshold $\nu_1$ and (iii) class III, when their MSF is negative inside an interval bounded by $\nu_1$ and $\nu_2$. This classification can be adapted to layers, just by considering each of them as a unique dynamical system with $N \times m$ dimensions. This way, we could define a class I,II or III layer according to the synchronization properties previously defined in the framework of dynamical systems}, thus preventing the occurrence of intra-layer synchronization for any possible value of $\sigma$ at $\lambda=0$. In addition, the inter-layer coupling function is taken to be ${\bf H}=(0,y,0)$, which generates a class II MSF at $\sigma=0$.
Results are reported in Fig.~\ref{fig2}, where $E_{inter}$ is plotted versus $\lambda$ for several values of  $\sigma$, both for SF (a) and ER (b)
topologies. In all cases, a smooth transition from an incoherent multiplex dynamics with $E_{inter}>0$ to an inter-layer synchronous evolution where $E_{inter}=0$ is observed, always in the absence of intra-layer synchronization [insets in Fig. \ref{fig2}(a,b) show that $E_{intra}$ remains well above zero for the whole explored range of $\lambda$]. In Fig.~\ref{fig2} (c) the MLE for the SF case is plotted, showing that $E_{inter}$ vanishes exactly at the same $\lambda$ at which the MLE gets negative, thus confirming the validity of the analytical approach. To gather a clearer view on the impact of the network heterogeneity, Fig.~\ref{fig2} (d) reports the critical coupling $\lambda^*$ (the value of $\lambda$ at the onset of inter-layer synchronization) as a function of $\sigma$,
for both SF and ER topologies, and several average degrees. As  in single layer networks, multiplexes of heterogeneous structures require smaller coupling thresholds to sustain a stable synchronous state. There is a non-monotonic relationship between the synchronization threshold and the
{\it stiffness}
within each layer (as measured by $\sigma$). The horizontal (vertical) dashed line in Fig.~\ref{fig2}(d) [Fig.~\ref{fig2}(a,b)] indicates the threshold $\lambda^*$ for the appearance of a synchronous state at $\sigma=0$, obtained by analyzing a pair of bidirectionally coupled R\"ossler systems. More rigid layers need larger inter-layer couplings to synchronize (as one would expect), but beyond a certain point in the rigidity, the trend is remarkably reversed.

\begin{figure}
  \centering
  \centering\includegraphics[width=\linewidth]{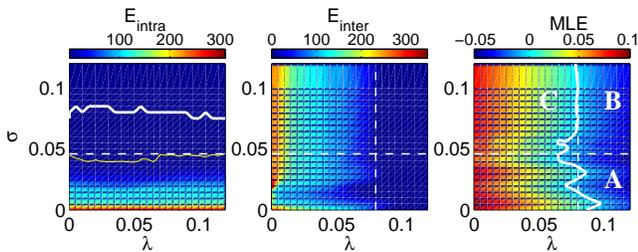}
  \caption[]{(Color online). 
The intra- (left) and inter-layer (middle) synchronization errors (see main text for definitions), and the MLE (right) in the ($\sigma$, $\lambda$) parameter space. Corresponding color codes are shown in the upper bars.
In all panels, the horizontal dashed lines mark the synchronization  threshold of each isolated layer ($\lambda=0$). In the left panel, the vertical dashed line marks the synchronization threshold of a pair of nodes ($\sigma=0$). The thick white (thin yellow) contour line in the left panel is the isoline corresponding
to $E_{intra}=0$ ($E_{intra}=0.01 E_{intra}^{max}$). The thick white contour line in the right panel is the isoline where the MLE changes its sign from positive to negative. See main text for the description of regions A, B and C in the right panel.
Each point is an average over 10 multiplexes realizations with $\langle k\rangle=16$.  \label{fig3}}
\end{figure}

A much richer scenario occurs in the case ${\bf h}=(x,0,0)$, where the uncoupled layers ($\lambda=0$)
are of class III \cite{Note2}, and therefore inter- and intra-layer synchronization can, in principle, coexist. We make use of an ER multiplex network of $\langle k\rangle=16$ to show the interplay of both types of synchronization \footnote{Similar qualitative results, not shown here, are obtained with different average degrees and SF structures.}. The results are reported in Fig.~\ref{fig3}.
In particular, the left panel shows that $E_{intra}$ vanishes (thick white contour line) for values of $\sigma$ slightly above the one predicted by the MSF for intra-layer synchronization at $\lambda=0$ (white dashed line). However, assuming a margin of error of $1\%$, approximate intra-layer synchronization (thin yellow line) can be reached for values of $\sigma$ even smaller than the case in which nodes are not multiplexed.
This shows how intra-layer synchronization is only mildly affected by the presence of inter-layer couplings.
The middle and right panels of  Fig. \ref{fig3} report $E_{inter}$ and the MLE, respectively.
In both panels, the vertical dashed line further marks the synchronization transition point predicted by the MSF for two coupled oscillators ($\sigma=0$). Three different regions (A,B, and C) can be identified in the parameter space:  inter-layer without (region A) and with (region B) intra-layer synchronization, and an area
(region C) where intra-layer synchronization occurs without inter-layer synchronization. In the right panel, the isoline (white thick curve) marks the points where the MLE changes its sign from positive to negative, and shows that, at intermediate values of $\sigma$, inter-layer synchronization is realized for values of $\lambda$ below the synchronization threshold
of a pair of oscillators (vertical dashed line). A second remarkable conclusion is that, in a multiplexed structure, inter- and intra-layer synchronization may enhance each other.
\begin{figure}
    \centering\includegraphics[width=\linewidth]{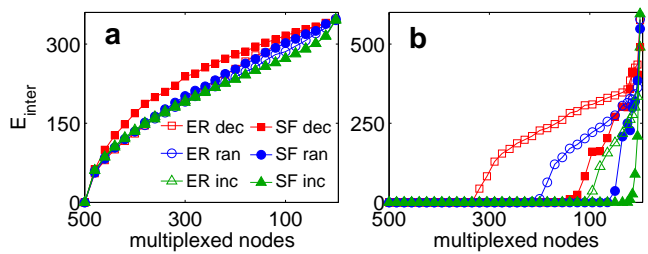}
  \caption[]{(Color online). $E_{inter}$ vs. the number of multiplexed nodes for ER (void symbols) and SF (solid symbols) configurations. From a full multiplex, nodes are progressively disconnected following a random  (blue circles), and a decreasing (red squares) or increasing degree (teal triangles) sequence. $\lambda=0.1$ (a) $\sigma=0.1$, (b) $\sigma=1.0$. Points are averages over 20 network realizations, with  $\langle k\rangle=8$.  \label{fig4}}
\end{figure}
\begin{figure}
  \centering\includegraphics[width=\linewidth]{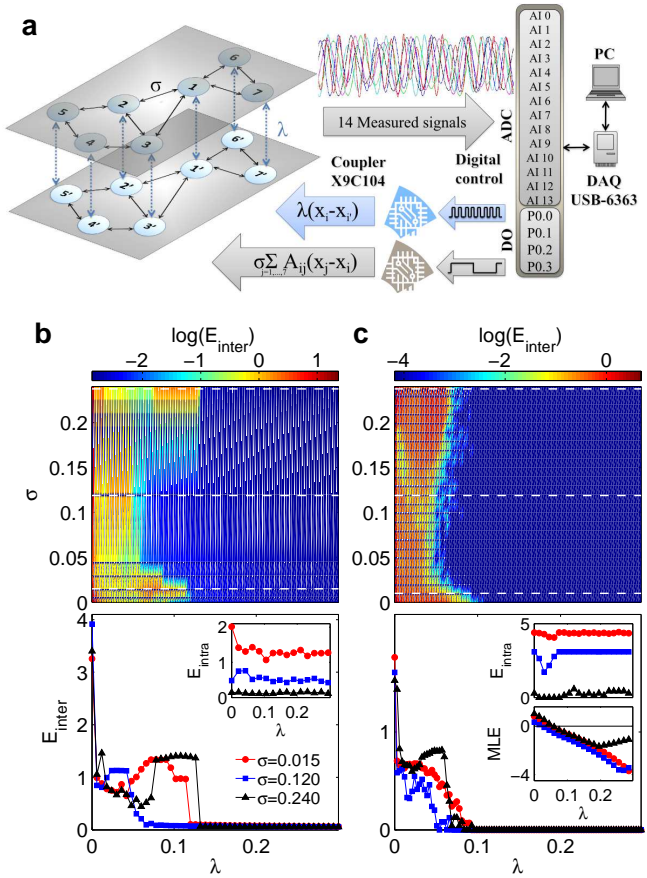}
  \caption[]{(Color online). (a) Experimental set-up. The left image is a sketch of the coupling topology of the 14 electronic circuits
composing the multiplex network
(see main text for the description of the experimental procedure used). The whole experiment is controlled from a PC with Labview Software. (b-c) Color maps of $E_{inter}$ (log scale) in the parameter space $(\sigma,\lambda)$ (top panels) and for three specific $\sigma$ values (bottom panels, color codes in the legend) calculated experimentally (b) and via numerical simulations (c). Insets show the corresponding values of $E_{intra}$. In panel c, the MLE is also reported as a separate inset.
\label{fig5}}
\end{figure}

Further insight can be gathered by exploring the robustness of the inter-layer synchronous state under a progressive de-multiplexing of the structure. For both the SF and ER architectures and starting from the complete multiplex, we then sequentially remove the links between nodes and their corresponding replicas, until the two layers become uncoupled. In Fig.~\ref{fig4}, $E_{inter}$ is reported as a function of the actual number of multiplexed nodes, from $N$ to $0$, with a disconnecting mechanism following either a random sequence or the increasing/decreasing degree ranking.
Robustness is critically dependent on the balance between the inter- and intra-layer couplings. at relatively low and balanced couplings (left panel) $E_{inter}$ grows as soon as the first pair of replica nodes is disconnected, and almost at the same rate regardless on the node sequence. A radically different situation occurs when the intra-layer coupling considerably exceeds the inter-layer one (right panel): inter-layer synchronization persists even if a large fraction of nodes are de-multiplexed. Furthermore, multiplexes with homogeneous structured layers (void symbols) are less robust than those formed by SF layers (solid symbols), and engineering a multiplex with synchronous layers is actually tantamount to coupling just a fraction of the largest degree nodes in each layer. This behavior holds even when the hubs of the SF multiplex are sequentially disconnected [see squares of Fig. \ref{fig4}(b)]. Notice indeed that, in analogy with what reported for network's targeting \cite{Gutierrez2012}, only $25$ ($110$) of the largest degree nodes maintain $E_{inter}=0$ in SF (ER) multiplexes of size $N=500$.

Finally, we report experimental evidence of inter-layer synchronization in nonlinear electronic circuits, with the setup sketched in Fig.~\ref{fig5}(a). The experiment consists of an electronic array,
a personal computer (PC), 14 analog to digital converters (ADC) and 4 digital ports (DO) from a multifunctional data card (DAQ)
controlled by Labview. The ADC's are used for sampling one of the state variable out of all the networked circuits,
the DO's are used as controllers for the gain of the two coupling strengths $\sigma$ and $\lambda$.
The array is made of 14 R\"ossler-like circuits arranged in two identical layers (blue nodes), each one of them having two different
electronic couplers, one for the coupling among nodes in the same layer ($\sigma$) and the second for the interaction of each node
with its replica in the other layer ($\lambda$). The chaotic dynamics of the circuits is well approximated by ${\bf f}({\bf x})=[-\alpha_1(x+\beta y +\Gamma z),-\alpha_2(-\gamma x + (1-\delta)y),-\alpha_3(-g(x) +z)]$, with $g(x)=0$ if $x\le 3$ and $g(x)=\mu(x-3)$ if $x> 3$ (see \cite{Sevilla2015SR} for a detailed description of the experimental implementation of the R\"ossler like circuit in networks, and \cite{Aguirre2014,Sevilla2015Le} for previous realizations in different network configurations).
The coupling is adjusted using two digital potentiometers X9C104, whose parameters $C_{u/d}$ (Up/~Down resistance) and $C_{step}$ (increment of the resistance at each step) are controlled by digital signals coming from a DAQ Card, P0.0-P0.3. The outputs of the circuit are sent to a set of voltage followers that act as a buffer, and then  to the analog ports (AI0-AI13) of the same DAQ Card.
At each $\sigma$ value (starting from $\sigma=0$), $\lambda$ is initially set to zero, and then the polarization voltage of the circuits is turned off and on, after a waiting time of 500 ms. The signals corresponding to the $x$ state variables of the 14 circuits are acquired by the analog ports AI0-AI13 and saved in the PC for further analysis.  $\lambda$ is then incremented by one step, and the procedure is repeated 100 times (until the maximum value of $\lambda$ is reached). When the entire run is finished, $\sigma$ is increased by one step, and another cycle of $\lambda$ values is initiated. The experimental (panel b) and numerical (panel c) results for $E_{inter}$ and $E_{intra}$ are in very good agreement for the entire parameter space $(\sigma,\lambda)$, indicating that our analytical predictions actually apply also for slightly non identical systems, as the electronic circuits contain resistors and capacitors of 1\% and 10\% tolerance, respectively, causing a small deviation with respect to the synchronization region predicted by the MSF approach \cite{Sevilla2015Le}.

In conclusion, we provided a full characterization of inter-layer synchronization,
a novel and distinctive dynamical phenomenon occurring in multiplex networks, in terms of its stability conditions, its relation to intra-layer synchronization and network topology, and its robustness under partial de-multiplexing of the network. We further reproduced it experimentally  for slightly non identical systems, indicating that the phenomenon is robust enough to be observable in the presence of noise and parameter mismatch.
Our results, therefore, suggest the way of unveiling the new dynamics in a variety of  multiplexed real world systems.

Work partly supported by the Ministerio de Econom\'ia y Competitividad
of Spain under projects FIS2012-38949-C03-01 and FIS2013-41057-P.
ISN and IL acknowledge support from GARECOM, Group of Research Excelence URJC-Banco de Santander.
The Authors also acknowledge the computational resources and assistance provided by CRESCO,
the supercomputacional center of ENEA in Portici, Italy. 
R.S.E. acknowledges Universidad de Guadalajara, CULagos (Mexico) for financial support (PRO-SNI/228069, PROINPEPRG/005/2014,
UDG-CONACyT/I010/163/2014).

\bibliography{references}
\end{document}